\input{aipcheck}

\documentclass[
    ,final            
  ]
  {aipproc}

\layoutstyle{6x9}

\newcommand{\rmd}{\mathrm{d}}

\newcommand{\be}{\begin{equation}} 
\newcommand{\ee}{\end{equation}} 
\newcommand{\Be}{\begin{eqnarray*}}
\newcommand{\Ee}{\end{eqnarray*}}
\newcommand{\bey}{\begin{eqnarray}} 
\newcommand{\eey}{\end{eqnarray}} 

\newcommand{\lb}{\label}

\begin{document}

\title{The Hamiltonian Mean Field model: anomalous or normal diffusion?}
 

\author{Andrea Antoniazzi}{
  address={Dipartimento di Energetica ``S. Stecco" and CSDC, Universit\`a di
  Firenze, and INFN, via S. Marta, 3, 50139 Firenze, Italy}
}

\author{Duccio Fanelli}{
  address={Theoretical Physics, School of Physics and Astronomy, University of Manchester, 
  Manchester M13 9PL, United Kingdom}
}
   
\author{Stefano Ruffo}{
  address={Dipartimento di Energetica ``S. Stecco" and CSDC, Universit\`a di
  Firenze, and INFN, via S. Marta, 3, 50139 Firenze, Italy}
}

\begin{abstract}
We consider the out-of-equilibrium dynamics of the Hamiltonian Mean Field (HMF) model, 
by focusing in particular on the properties of single-particle diffusion. As we shall here 
demonstrate analytically, {\it if} the autocorrelation of momenta in the so-called 
quasi-stationary states can be fitted by a $q$-exponential, {\it then}
diffusion ought to be normal for $q<2$, at variance with the interpretation of 
the numerical experiments proposed in Ref.~\cite{rapisarda}.
\end{abstract}

\maketitle

\section{Introduction}
Several numerical studies have revealed that systems with long-range interactions present many 
interesting, and pretty unique, dynamical features which take place out-of-equilibrium, 
manifesting a strong sensitivity to initial conditions.
To investigate these peculiar aspects and eventually develop a comprehensive theoretical framework, 
it is extremely valuable to dispose of a simple toy-model that admits   
an Hamiltonian formulation in terms of continuous variables. A paradigmatic example is represented by the so-called  
Hamiltonian Mean Field (HMF) model \cite{antoni-95}, which describes the evolution of $N$ particles 
coupled through an equally strong attractive cosine interaction. 
The model is specified by the following Hamiltonian:
\begin{equation}
\label{eq_ham}
H = \frac{1}{2} \sum_{j=1}^N p_j^2 + \frac{1}{2 N} \sum_{i,j=1}^N [1 - \cos(\theta_i-\theta_j)]
\end{equation}
where $\theta_i \in [0,2\pi[$ represents the position (angle) of the $j$-th particle on a unitary 
radius circle and $p_j$ stands for its conjugate momentum. The HMF model can also be seen as a simplification of the gravitational 
sheet model \cite{Feix}, when considering only the first harmonic in the Fourier expansion of the potential. It 
is exactly solvable both in the canonical and microcanonical ensembles, leading in 
this case to equivalent results, and displays a second order phase transitions from a
homogeneous to a clustered phase when decreasing energy or temperature.
The most striking observations relate
however to nonequilibrium dynamics. Indeed, when performing numerical simulations starting out-of-equilibrium, 
the system remains trapped in long-lived Quasi Stationary States (QSSs), before relaxing to thermodynamic equilibrium.  
In this dynamical regime the evolution is particularly slow and the system displays non-Gaussian momentum 
distributions\cite{rapisarda,antoniazzi_prl}. To monitor the evolution of
the system, it is customary to introduce the magnetization, 
an order parameter defined as $M=|{\mathbf M}|=|\sum {\mathbf m_i}|
/N$, where ${\mathbf m_i}=(\cos \theta_i,\sin \theta_i)$ is the
magnetization vector. Simulations are often carried out for 
water-bag initial distributions, for which the one-body distribution function $f(\theta,p,t)$
takes at $t=0$ a non vanishing constant value only inside the rectangular phase-space domain $D$ specified by
\begin{equation}
D = \{(\theta,p)\in [-\pi,\pi] \times [-\infty,\infty]~|~ |\theta|<\Delta\theta, ~|p|<\Delta p\},
\end{equation}
where $0\leq\Delta\theta\leq\pi$ and $\Delta p \geq 0$. The initial magnetization $M_{0}$ and 
the energy density $U$ can be expressed as functions of $\Delta\theta$ and $\Delta p$
\begin{displaymath}
M_{0} = \frac{\sin(\Delta\theta)}{\Delta\theta} ,
\quad
U = \frac{(\Delta p)^{2}}{6} + \frac{1-(M_{0})^{2}}{2}~.
\end{displaymath}
This in turn implies that the initial water-bag profiles are uniquely determined by $M_{0}$ and $U$,
which take values in the ranges $0 \leq M_{0} \leq 1$ and $U \geq (1 - M_0^2)/2$.

The mean square angular displacement $\sigma^2(t)=\frac{1}{N}\sum_i{[\theta_i(t) -
\theta_i(0)]^2}$ is also a quantity of interest. The scaling
$\sigma^2 \propto t^{\gamma}$ defines the diffusive behavior:
$\gamma=1$ corresponds to normal diffusion and $\gamma=2$ to free
particle ballistic dynamics.  Intermediate cases correspond to the
anomalous {\it superdiffusive} behavior. On the numerical simulations side, it is claimed in
Ref.~\cite{rapisarda} that QSSs display anomalous diffusion
with an exponent $\gamma$ in the range $1.4-1.5$ for $0.4 \le M_0 \le 1$ 

The same authors investigated the decay of the momentum 
autocorrelation functions and suggested that the numerical profiles are well interpolated by 
a so-called $q$-exponential function \cite{tsallis}. Based on this finding,  
a possible relation between the fitted value $q$ and the diffusion exponent $\gamma$ has been 
hypothesised  \cite{rapisarda}. 
 
We shall here analytically prove that the aforementioned results are contradictory. 
More specifically, assuming a $q$-exponential decay of the autocorrelation of momenta {\it implies}
normal diffusion, for the range of values of $q$  reported in Ref.~\cite{rapisarda}. 

Angular diffusion and momentum autocorrelation for the HMF model have been often discussed 
in the literature \cite{antoniazzi_prl, latora_rapisarda_ruffo, YAMA, thierry}. 
In order to provide a comprehensive picture on this subject, we will 
shortly review some of these contributions in the next Section. 
The following Section is instead devoted to presenting our analytic argument. Finally       
we sum up and draw our conclusions.

\section{On the relation between relaxation and diffusion for the HMF model}
\label{sub_diff_review}

Numerical evidence for a superdiffusive behaviour in the HMF model was first reported in 
\cite{latora_rapisarda_ruffo}. In this paper a water-bag initial condition \footnote{In the papers hereafter discussed, 
 the choice $U=0.69$ is always put forward, when a water--bag profile is initially assumed.} was assumed, with the particles 
positioned in $\theta_i=0$, a choice which corresponds to select initially the magnetized state $M_0=1$. 
It was also claimed in \cite{latora_rapisarda_ruffo} that the anomalous diffusion occurs for a transient
out--of--equilibrium regime, when the system is trapped in a QSS. 

Successive numerical experiments \cite{YAMA} carried on for a homogeneous initial distribution, i.e. $M_0=0$, 
revealed that  the anomalies in diffusion are instead associated to the (nonstationary) relaxation to equilibrium.  
The QSS should therefore be characterized by a normal diffusive regime: it was in fact argued that the 
anomalous diffusion detected in \cite{latora_rapisarda_ruffo} stems from finite size effects. 

Motivated by such controversial findings, the authors in \cite{rapisarda} set down to elucidate the role of initial conditions in
driving HMF dynamical anomalies. To this end the usual water--bag initial distribution was considered, with 
the particles confined in a variable portion of the unitary circle, so to span the whole interval $0 \le M_0 \le 1$. 
First, the momentum autocorrelation function was introduced as:
\begin{equation}
\label{Cq}
C_p(t;0) = \langle p(t)p(0) \rangle_N = \frac{1}{N}\sum_i^N p_i(t)p_i(0),
\end{equation}
and its decay monitored, for different values of $M_0$. The numerical curves were then interpolated with the 
$q$-exponential:
\begin{equation}
\label{eq}
e_q(t,\tau)=[1-(1-q)t/\tau]^{\frac{1}{1-q}},
\end{equation}
a function which arises in the realm of Tsallis' generalized thermostatistics \cite{tsallis}. 
Both $\tau$ and $q$ are numerically adjusted in \cite{rapisarda}: $q$ is empirically found to be $1.5$ 
for $M_0 \ge 0.4$, while it progressively diminuishes for smaller values of $M_0$. For $M_0=0$ an 
almost exponential decay is measured. In this respect, the homogeneous condition was assigned a 
special nature, and claimed to be intrinsically different from the initially magnetized configurations. 
This conclusion was further strengthened by looking at the diffusive properties of the system. 
For $M_0>0$, superdiffusion is systematically observed in Ref.~\cite{rapisarda}, while the case $M_0=0$ 
tends to a normal diffusive behavior when increasing the number of simulated particles, as previously found in 
\cite{YAMA}. Moreover, a closed analytical relation was conjectured 
between the index $q$ and the diffusion exponent $\gamma$, based on an generalization
of the standard Fokker-Planck equation \cite{tsa-buk}. Such relation reads \cite{rapisarda,tsa-buk}:
\be\lb{q_of_gamma}
\gamma = \frac{2}{3-q}
\ee
and returns values of $\gamma$ similar to those 
measured in \cite{rapisarda}, when the corresponding $q$ are inserted. This supposed agreement 
is merely a coincidence, as we shall prove in the following. If one assumes a $q$-exponential decay 
for the autocorrelation of momenta, then normal diffusion is to be exptected, for the range of $q$  
reported in \cite{rapisarda}. Before turning to illustrate this point, we shall devote the remaining 
part of this Section to 
complete our review on the studies of diffusion for the HMF model.

Among other interesting contributions, Ref.~\cite{anteneodo} is worth mentioning: here the controversy on the role of the initial
condition is also addressed. In particular a fully magnetized ($M_0=1$) initial condition is investigated. For finite sizes, a superdiffusive
phase is indeed detected, in agreement with \cite{rapisarda,latora_rapisarda_ruffo}. However, the local exponent $\gamma$ is shown to
converge to the normal diffusion value $\gamma=1$ as the numbers of particle is increased. Thus, anomalous diffusion is just a finite size effect, 
an observation which confirms the scenario proposed in \cite{YAMA} for the homogeneous initial setting. A similar conclusion is also
reached in \cite{antoniazzi_prl}, where the time evolution of $\sigma^2$ is monitored by employing a larger number of particles than in
previous studies. An almost normal diffusion is reported also for intermediate values of the initial magnetization. 

Finally, in Ref.~\cite{thierry} a kinetic approach that goes beyond the Vlasov approximation was developed. This methodology
enables to esplicitly calculate the momentum autocorrelation function 
in terms of the momentum distribution at $t=0$, namely $f_0(p)$. This approach applies to homogeneous 
initial distributions and assumes that the system is initially in a QSS. The distribution $f_0(p)$
with algebraic tails corresponds to a correlation function of momenta with an algebraic 
decay in the long time regime, hence resulting in a superdiffusive behaviour. 
In contrast, for distributions with Gaussian tails, $C_p(t;0)$ scales as $\ln t/t$, which in turn implies an
approximately normal scaling for the mean--square displacement of the angles. The validity of these predictions has been recently
confirmed in \cite{yama_jstat}, where a detailed campaign of simulations was performed. It is however worth emphasising that the theory
outlined in \cite{thierry} is solely limited to homogeneous initial conditions. Furthermore, since the theory states that 
the asymptotic law of diffusion is determined by the tails of the distribution, it does not directly apply to the 
case where a water--bag initial condition for the momenta is selected. 
The latter has in fact no tails initially, although Gaussian tails rapidly develop in the course of time, as shown in  
\cite{antoniazzi_prl}, thus possibly pointing to a normal diffusion.

\section{$q$--exponential decay of momentum autocorrelation implies normal diffusive behaviour}
\label{sub_diff}

As anticipated, it is assumed in Ref.~\cite{rapisarda} that the decay of the momentum autocorrelation function can be 
fitted by a $q$-exponential function \cite{rapisarda}. These results are 
recurrently invoked to support the claim of anomalous diffusion, the exponent $\gamma$ and the 
index $q$  being related through expression (\ref{q_of_gamma}). As opposite to this working ansatz, we will here analytically 
demonstrate that a $q$-exponential decay of the autocorrelation of momenta leads to normal diffusion, 
for the values of $q$ that have been fitted in \cite{rapisarda}. 

We start by noticing that
the quantity $\sigma^{2}(t)$ can be rewritten in terms of 
the correlation function of momenta $C_{p}(t;w)$ through an {\it exact} transformation \cite{YAMA,thierry}:
\bey
  \label{eq:variance2}
    \sigma^{2}(t) 
    && = \int_{0}^{t} \rmd t_{1} \int_{0}^{t} \rmd t_{2}~ 
    \langle p_{j}(t_{1})p_{j}(t_{2})\rangle_{N} \\
    && = 2 \int_{0}^{t} \rmd s \int_{0}^{t-s} \rmd w~
    C_{p}(s;w),
\eey
where $C_{p}(t;w)$ is defined as
\begin{equation}
  \label{eq:corrp}
  C_{p}(t;w) = \langle p_{j}(t+w) p_{j}(w) \rangle_{N}.
\end{equation}
Moreover, if the system is stationary, 
$C_{p}(t;w)$ does not depend on $w$ and hence
\begin{equation}
	  \label{eq:stationarity}
  C_{p}(t;w)=C_{p}(t;0)  \quad (\forall w>0)
\end{equation}
then Eq.(\ref{eq:variance2}) is simplified into
\begin{equation}
  \label{eq:simplified-variance2}
  \sigma^{2}(t) = 2 \int_{0}^{t}
  (t-s)~C_{p}(s;0)~\rmd s.  
\end{equation}
Now, let us assume a $q$-exponential decay for $C_p(t;0)$, namely:
\begin{equation}
\label{Cq1}
C_p(t;0) = [1-(1-q)t/\tau]^{\frac{1}{1-q}},
\end{equation}
and insert the expression for $C_{p}(s;0)$ into (\ref{eq:simplified-variance2})
\bey
\label{diff_2}
\sigma^2(t) &=& 2\int_0^t (t-s)[1-(1-q)s/\tau]^{\frac{1}{1-q}}ds  \\
&=&2\int_0^t t[1-(1-q)s/\tau]^{\frac{1}{1-q}}ds - 
2\int_0^t s[1-(1-q)s/\tau]^{\frac{1}{1-q}}ds.\nonumber
\eey
The first term on the right hand side of the previous equation gives:
\bey
\label{diff_3}
\int_0^t t[1-(1-q)s/\tau]^{\frac{1}{1-q}}ds &=& t  \frac{\tau}{1-q}\int_0^{\frac{t(1-q)}{\tau}} 
[1-z]^{\frac{1}{1-q}}dz =\\
&=& -t  \frac{\tau}{2-q} 
[1-t(1-q)/\tau]^{\frac{2-q}{1-q}}  + t  \frac{\tau}{2-q},\nonumber
\eey
while the second yields: 
\bey
\label{diff_4}
\int_0^t &&s[1-(1-q)s/\tau]^{\frac{1}{1-q}}ds =
-t  \frac{\tau}{2-q} [1-t(1-q)/\tau]^{\frac{2-q}{1-q}} +\\
&& \quad\quad\quad -\frac{\tau^2}{(2-q)(3-2q)} [1-t(1-q)/\tau]^{\frac{3-2q}{1-q}} 
+ \frac{\tau^2}{(2-q)(3-2q)}. \nonumber
\eey
Finally, collecting together the two contributions, $\sigma^2$ takes the form:
\begin{equation}
\label{diff_5}
\sigma^2(t) =  2t \frac{\tau}{2-q}
 + 2\frac{\tau^2}{(2-q)(3-2q)} [1-t(1-q)/\tau]^{\frac{3-2q}{1-q}} 
- 2\frac{\tau^2}{(2-q)(3-2q)}.
\end{equation}
The first term corresponds to a normal diffusion ($\gamma=1$), while the latter is constant
and can be therefore neglected for long enough times. More importantly the second term results in 
a  contribution which is proportional
to $(1 \pm Ct)^{\nu}$ ($-$ if $q<1$ and $+$ if $q>1$), where $\nu=(3-2q)(1-q)$.
Now, it is clear that this term causes the diffusion to be anomalous only if $q>2$, otherwise
resulting in a correction to the $\sigma^2 \propto t$ behaviour.

\begin{figure}
\includegraphics[width=10cm]{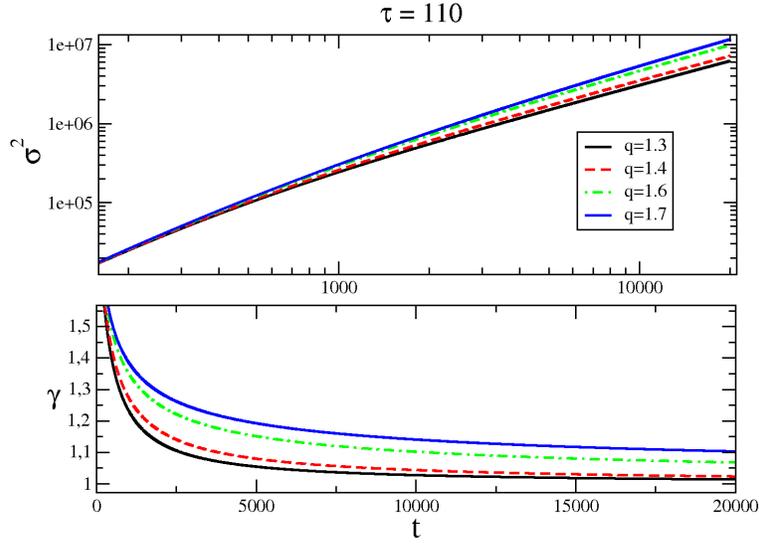}
\caption{Upper panel: Time evolution of the angular diffusion as predicted by Eq. (\ref{diff_5}) for different values
of $q$. Lower panel: Time evolution of the coefficient $\gamma$, evaluated through the logarithmic derivative 
\label{fig_diffusione}.}
\end{figure}

The functions (\ref{diff_5}) have been plotted in the upper panel of Fig~\ref{fig_diffusione} for different values
of $q < 2$. In the lower panel we have reported, for the same functions, the evolution of the diffusion
coefficient $\gamma$, which are is consistent with the results in \cite{antoniazzi_prl,YAMA}.

\section{Conclusion} 
\label{sub_concl}

In this paper we have reviewed the interlaced connection between relaxational and diffusional properties of the HMF
model. 
This is a debated issue  which has led to highly controversial interpretations in the recent times. 
In this respect, we have here critically discussed the interpretation of the  
data presented in Ref.~\cite{rapisarda}.
Our main conclusion goes as follows: 
having found a momentum autocorrelation function which is correctly interpolated by the relation (\ref{Cq1}), with the choice 
$q \sim 1.5$~\cite{rapisarda}, should necessarily lead to normal diffusion (asymptotic linear growth
in time of the second moment). 
The claim of anomalous diffusion put forward in the same series of papers~\cite{rapisarda} seems 
therefore in contradiction with the proposed fit of the autocorrelation function.

Indeed, such apparent contradiction could be due to finite size effects as pointed out in \cite{YAMA,anteneodo}. 
Considering that the asymptotic convergence is very slow  \cite{YAMA, thierry, anteneodo}, the estimates for 
the anomalous exponent are  probably affected by 
the limited number of  particles, an unfortunate fact which acts as a bias towards values of $\gamma$ larger than $1$. 
Moreover, the time required to converge to the normal diffusion regime is remarkably long \cite{antoniazzi_prl,thierry}.
It is hence also possible that the time window explored by the authors in their investigations~\cite{rapisarda} 
is too short. 

\end{document}